# Silicon Nanoantenna Mix Arrays for a Trifecta of Quantum Emitter Enhancements

Zhaogang Dong[1,*], Sergey Gorelik[1], Ramón Paniagua-Dominguez[1], Johnathan Yik[2], Jinfa Ho[1], Febiana Tjiptoharsono[1], Emmanuel Lassalle[1], Soroosh Daqiqeh Rezaei[3], Darren C. J. Neo[1], Ping Bai[2], Arseniy I. Kuznetsov[1,*], and Joel K. W. Yang[3,1,*]

[1]Institute of Materials Research and Engineering, A*STAR (Agency for Science, Technology and Research), 2 Fusionopolis Way, #08-03 Innovis, 138634 Singapore

[2]Institute of High Performance Computing, A*STAR (Agency for Science, Technology and Research), 1 Fusionopolis Way, #16-16 Connexis, 138632 Singapore

[3]Singapore University of Technology and Design, 8 Somapah Road, 487372, Singapore

*Correspondence and requests for materials should be addressed to J.K.W.Y. (email: joel_yang@sutd.edu.sg), A.I.K. (email: Arseniy_Kuznetsov@imre.a-star.edu.sg) and Z.D. (email: dongz@imre.a-star.edu.sg).

**ABSTRACT**

Dielectric nanostructures have demonstrated optical antenna effects due to Mie resonances. Preliminary investigations on dielectric nanoantennas have been carried out for a trifecta of enhancements, *i.e.,* simultaneous enhancements in absorption, emission directionality and radiative decay rates of quantum emitters. However, these investigations are limited by fragile substrates or low Purcell factor, which is extremely important for exciting quantum emitters electrically. In this paper, we present a Si mix antenna array to achieve the trifecta enhancement of ~1200 fold with a Purcell factor of ~47. The antenna design incorporates ~10 nm gaps within which fluorescent molecules strongly absorb the pump laser energy through a resonant mode. In the emission process, the antenna array increases the radiative decay rates of the fluorescence molecules via Purcell effect and provides directional emission through a separate mode. This work could lead to novel CMOS compatible platforms for enhancing fluorescence for biological and chemical applications.

Silicon (Si) is the most widely used material in the semiconductor industry, and Si-based technology platforms have developed rapidly in the past decades. Due to its high refractive index and relatively low Ohmic losses in the visible spectrum, Si nanostructures exhibit localized Mie resonances.[1-10] For instance, Si nanostructures support both electric and magnetic dipole resonances,[1,2] where the interactions of these dipoles lead to directional scattering.[11,12] In addition, dielectric nanostructures with Mie resonances, so-called dielectric nanoantennas, have enabled various optical applications, such as color printing,[5,13-15] negative-angle refraction,[16] nonlinearity enhancement,[17-20] spectrometers,[21] and optical holograms.[22-24]

Localized field enhancement is crucial for enhancing both absorptive and emissive processes in quantum emitters placed in these so-called hotspots. Despite the tendency for optical fields to be localized within the Si structures, instead of being confined to the surface as in plasmonic nanostructures, localized field enhancements are still achievable through closely spaced Si dimer nanostructures.[25] Though Si antennas have recently been shown to also exhibit interband plasmonic behavior with orders of magnitude field enhancements, this effect is observed only in the UV spectrum.[26] Nanogaps thus prove suitable for absorption and Purcell-factor enhancement[27] in both plasmonic and dielectric antennas, with dielectrics having an advantage of generating less heat.[28]

In addition to large field enhancements, quantum emission can be further enhanced through improved directivity of dielectric antennas.[29,30] Emission of fluorescent dye molecules deposited onto single Si dimer antennas is enhanced through hot spots within the gaps.[28,31] Other approaches include coupling of quantum dot emission to the localized Mie resonances of Si nanostructures,[9,32-36] grating waveguides,[37] asymmetric metasurfaces,[38] surface lattice resonances,[39] and hybrid dielectric-metal nanoantennas.[40, 41] In these works, photoluminescence (PL) enhancements are often achieved by aligning the localized Mie resonances either with the excitation or emission wavelength, or more rarely via directionality engineering. Although trifecta enhancements (*i.e.*, simultaneous enhancements in absorption, radiative decay rates and directionality) were demonstrated on a suspended membrane with hole array exhibiting photonic crystal effect, this design may have limitations in applications due to its fragile nature and difficulty in scaling.[42] In addition, preliminary trifecta enhancements are also explored via waveguide mode[37] and Mie resonance.[38] However, these achieved enhancement factors are limited to between ~8-100 fold[37, 38] with rather modest Purcell factors of ~3 fold[37, 38, 42] for radiative decay rate, which is the most important antenna characteristic for future electrical excitation of quantum emitters.

In this paper, we presented a Si mix antenna array with nanogaps of ~10 nm to achieve trifecta enhancements of ~1200 fold with a Purcell factor of ~47. First, we show that a simple Si square nanoantenna array with a side length of 140 nm and a gap size of 120 nm is able to achieve all the enhancement aspects at once (*i.e.,* enhanced absorption, directionality and radiative decay rates). Nevertheless, due to the relatively large gap size of 120 nm, this simple square nanoantenna array has a limited enhancement factor for absorption. To further improve the antenna performance, we then designed a mix nanoantenna array with a small gap of 12 nm to achieve a large PL enhancement of ~1200 fold experimentally. Simulations show that this enhancement arises from three effects, *i.e.*, enhanced absorption of ~25 fold, enhanced radiative decay rates of ~52 fold and enhanced directionality of ~2.45 fold. Our work on Si nanoantenna arrays highlight the importance of geometric design in producing a robust platform for enhancing fluorescence, e.g. for biological and chemical detection applications.[43-45]

Figure 1(a) shows the schematic of square Si nanoantenna array, where the Si used throughout this manuscript is single crystalline Si (c-Si) on sapphire substrate. Rhodamine 6G (R6G) molecules were deposited onto the antenna surface. We chose R6G molecules as the fluorescence dye for characterizing the optical performance of our nanoantenna array as R6G is a well-known bright fluorescent dye with a quantum yield of 95% in solution.[46] In this case, further enhancing the emission from this already-bright-dye with directionality control will prove useful in practical applications. The total emission enhancement factor $EF_{Total}$ due to c-Si nanoantenna array is the product of the following three factors:[47]

$$EF_{Total} = EF_{Abs} \times EF_{Rad} \times EF_{Direct}, \tag{1}$$

where $EF_{Abs}$, $EF_{Rad}$ and $EF_{Direct}$ denote the enhancement factors for pump laser absorption, radiative decay rates and emission directionality, respectively.

One method of enhancing the emission directionality is by engineering the lattice parameters of the nanoantenna array. Based on the Wood's Anomaly,[48, 49] the nanoantenna pitch is designed to be 260 nm (see details in Fig. S1). The c-Si nanoantenna has a height of 120 nm with a 10 nm residual c-Si layer to prevent charging during scanning electron microscopy (SEM) inspections. The antenna side length $L$ is designed to be 140 nm, so that its optical resonances are able to enhance the absorption of 532 nm pump laser as well as the radiative decay rate of R6G molecule emission. Figure 1(b) presents the SEM image of c-Si antenna array, with the fabrication process as shown in Fig. S2 based on inductively coupled plasma etching[50] and the measured refractive index in Fig. S3. Finally, R6G in ethanol with a concentration of 1 µmol/L was spin coated onto the sample surface at 800 revolutions-per-minute (rpm) for 1 min.

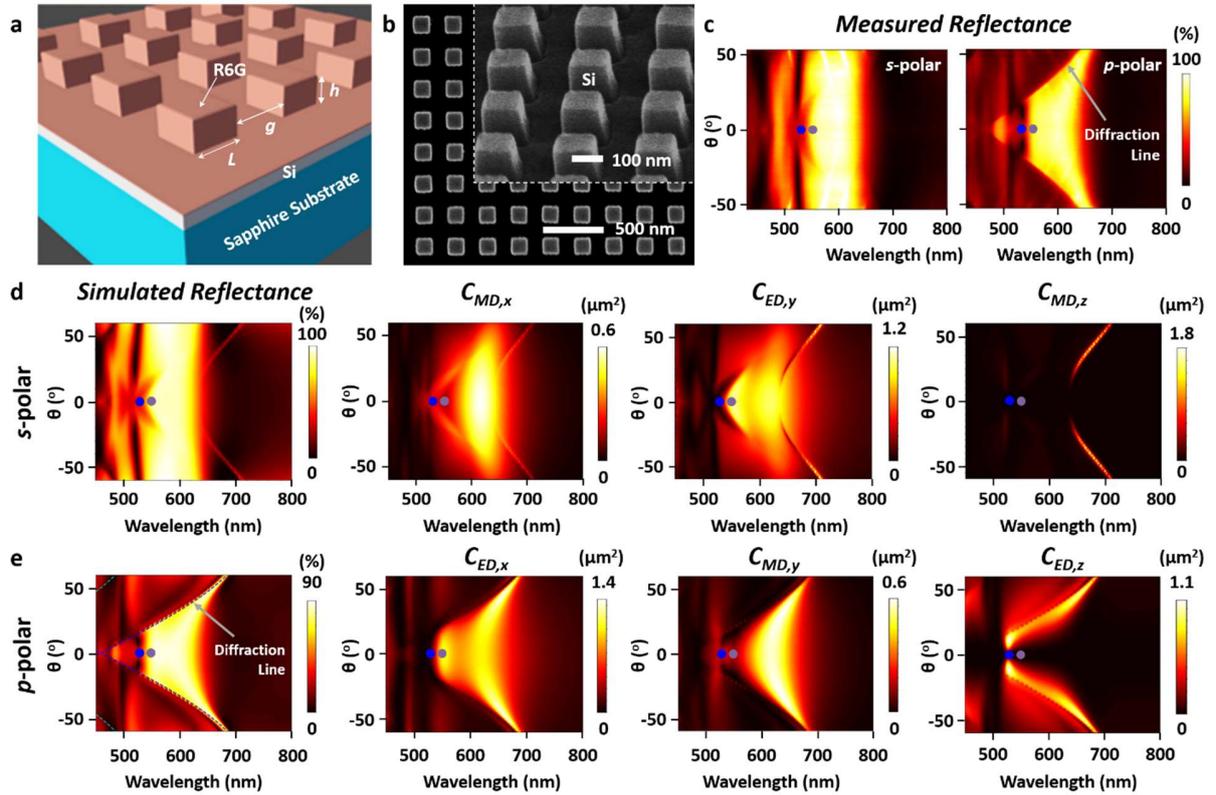

**Figure 1. Resonant optical modes of a square c-Si nanoantenna array.** (a)-(b) Schematic and scanning electron microscope (SEM) images of c-Si nanoantenna array with a side length *L* of 140 nm, a gap size *g* of 120 nm and a height *h* of 120 nm. (c) Measured angular reflectance spectra for two orthogonal polarizations. (d)-(e) Simulated angular reflectance spectra and multipolar decomposition. The pump laser wavelength and the R6G emission wavelength are denoted by blue and grey dots, respectively.

Figure 1(c) shows the measured angular reflectance spectra using back focal plane (BFP) spectroscopy.[51] The diffraction line in Fig. 1(c) is due to the 1$^{st}$ order Wood's Anomaly in the sapphire substrate, arising from the pitch size of 260 nm. Figures 1(d)-(e) present the simulated spectra and multipolar decompositions, where the electric dipole (*ED*) is the main component with the corresponding cross-sectional plots as shown in Figure S4. Here, this strong *ED* component is desirable to achieve the enhanced scattering for the PL emission from R6G molecules. In order to achieve the trifecta enhancements, we need to have structural resonances at both the pump and emission wavelengths of R6G molecules (as highlighted by blue and grey dots respectively), where the detailed discussions on respective enhancement mechanisms are shown in Fig. 2.

The 1st factor for PL enhancement is from the enhanced absorption of pump laser. Figure 2(a) presents the intrinsic absorption spectrum of R6G in ethanol, showing the peak absorptance at 530 nm. Wide-field PL images of the sample (*L*=140 nm & *g*=120 nm) with R6G molecules show excellent coating and fluorescence uniformity as shown in Fig. S5. The optimal pump wavelength of ~530 nm was determined from measuring PL intensity from the nanoantenna region as a function of $\lambda_{pump}$, as show in Fig. S6. This excitation spectrum clearly correlates with the intrinsic absorption spectrum of R6G molecule with the antenna resonances (Fig. 2). This peak absorption wavelength provides evidence that the R6G molecules has formed a monolayer coating, as the peak absorption wavelength would otherwise redshift from 530 nm to 560 nm if R6G film increases from monolayer to 10 layers.[52] Figure 2(a) also presents the normalized electric field distribution at $\lambda_{pump}$ of 532 nm, showing that this square antenna array already enhances the pump laser absorption by ~8 fold based on the overlap between the enhanced optical field distribution and R6G molecules.

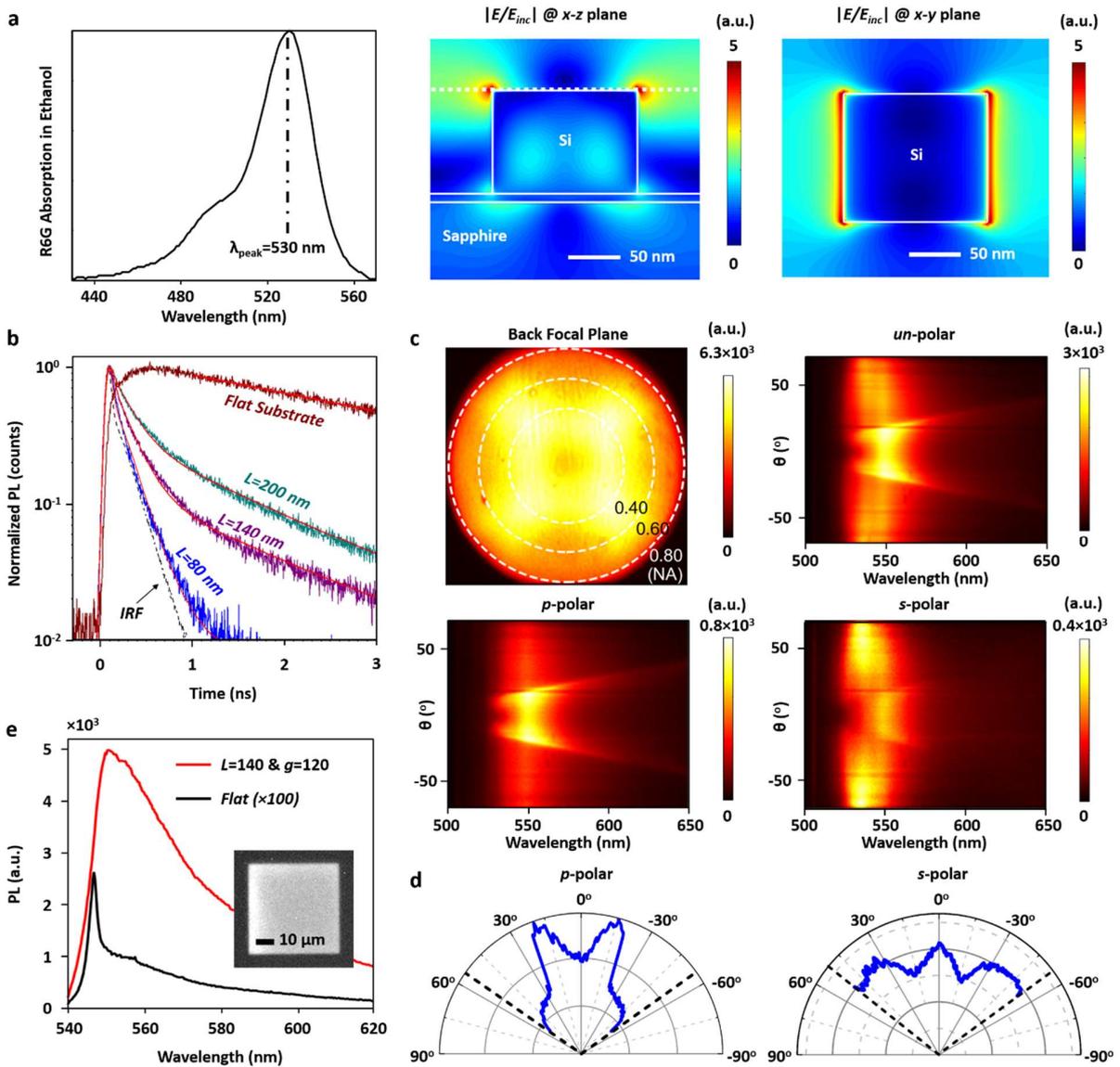

**Figure 2. Trifecta enhancement of directional emissions in basic square c-Si nanoantenna array.** (a) (left) Absorption spectrum of R6G molecules in ethanol. (right) Simulated electric field distribution $|E/E_{inc}|$ around c-Si nanoantenna at $\lambda_{pump}$ of 532 nm polarized along *x*-axis. (b) Radiative decay rate enhancement via lifetime measurements. *IRF* denotes the instrument response function. (c) PL directionality at back focal plane (BFP), measured PL spectra under un-polarized, *p*-polarized, and *s*-polarized conditions. (d) Polar plots of *s*- and *p*-polarized PL emission. (e) PL

spectra from the nanoantenna region and a flat region (×100 for better visibility) by 532 nm CW pump laser of showing a PL enhancement factor of ~450 fold. The inset figure is the PL image of the square antenna array with R6G molecules. All the PL emission spectra in (e) were measured from solid sample substrates.

The 2$^{nd}$ PL enhancement factor comes from enhanced radiative decay rates as characterized by emission lifetime measurements in Fig. 2(b). As reference, R6G emission from the flat region has a single exponential decay with a lifetime of ~3.2 ns, similar to its behavior in ethanol (see Fig. S7). This single exponential decay again indicates that the R6G film is a uniform monolayer without aggregates (such as dimers or excimers).[52] In contrast, the measured lifetime from the antenna ($L$=140 nm & $g$=120 nm) decreased to ~70 $p$s, indicating that the total decay rate (sum of radiative and non-radiative) is enhanced by ~46 fold, *i.e.,* Purcell factor $F_P$=~46. Notably, Fig. 2(b) also shows that nanoantennas with shorter side length $L$ increases decay rates; while gap size above 80 nm has little effect on the emission lifetime (see Fig. S8). These measurements show that R6G emission lifetime is predominantly dependent on the local Mie resonances of the individual nanoantennas. As derived in the supporting section S8, the Purcell factor $F_P$ is related to radiative decay rate and quantum yield by:

$$\frac{\gamma_r'}{\gamma_r} = F_P \times \frac{QY'}{QY}, \tag{2}$$

where $\gamma_r$ and $\gamma_r'$ denote the radiative decay rates from flat substrate and antenna region respectively. $QY$ and $QY'$ denote the quantum yields of the fluorescent molecules on the flat substrate and antenna region. Our simulations show that $QY$=~45.3% (see Fig. S9) and $QY'$=~40% (see Fig. S10). Hence, the enhanced radiative decay rate is determined to be ~41 fold.

The 3$^{rd}$ PL enhancement factor comes from the emission directionality. Figure 2(c)-(d) present the measured angular resolved PL spectra, where Fig. S11-S15 present the details of directionality simulations, near-field mode patterns and PL enhancement factor comparison. Based on the simulated PL directionality for the square c-Si nanoantenna array (see Fig. S13) and flat substrate (see Fig. S14), the enhanced directionality is determined to be ~2 fold. Therefore, based on Eq. (1), the overall PL enhancement factor is estimated to be ~656 fold. Experimentally, this square antenna array enhances the PL emission by ~450 fold as shown in Fig. 2(e).

Although the square nanoantenna array is able to enhance PL emission, it has not achieved the ultimate performance due to the relatively large gap size of 120 nm. For instance, smaller gaps are desired for better absorption since the optical field intensity in the gap increases exponentially with smaller dimensions.[25] As such, we introduce a mix antenna design with additional small square elements within the unit cell, as shown in Fig. S16. In addition, these smaller square elements can also lead to faster radiative decay rates as shown in Fig. 2(b). Therefore, the mix antenna design can potentially achieve the best of both enhanced absorption and radiative decay rates.

Figure 3(a) presents the schematic of mix antenna array, which has a unit cell with 4 basic square elements, *i.e.,* one square antenna element with a side length $L$ and three smaller square elements with a side length $L'$. With these dimensions, the mix nanoantenna will have multiple resonances, occurring at both the excitation and emission wavelengths. Here, the smallest gap size $g'$ between neighboring nanoantenna elements is designed to be ~12 nm so that intense optical fields are generated in the gap region, while such 12 nm gaps could be fabricated via electron beam lithography[53] or focused ion beam milling.[54] The pitch size is fixed at 260 nm for optimal directivity enhancement. Thus, the side length $L$, $L'$ and gap size $g'$ are related by:

$$L + L' + 2g' = 260 \ nm. \tag{3}$$

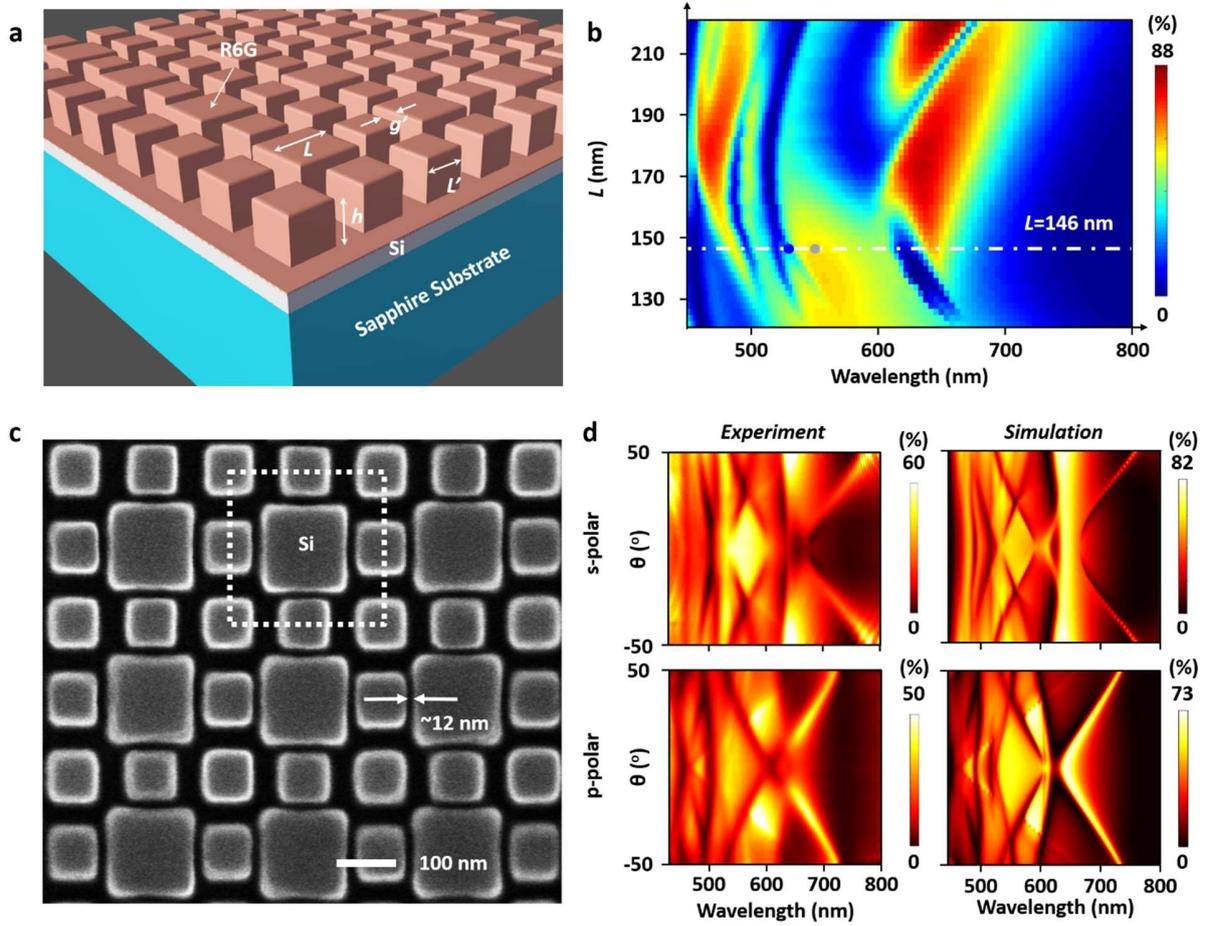

**Figure 3. Resonant optical modes of the mix nanoantenna array.** (a) Schematic of the mix nanoantenna array. Each unit cell consists of 4 basic square elements, *i.e.,* one square nanoantenna element with a side length *L* and three smaller square nanoantenna elements with a side length *L'*. The gap size *g'* is 12 nm. (b) Simulated reflectance spectra at normal incidence conditions (Γ point) with the side length *L* varied from 120 nm to 220 nm. (c) SEM image of the fabricated mix nanoantenna array with a minimum gap size of 12 nm, *L*=146 nm, and *L'*=90 nm. (d) Measured and simulated angular-resolved reflectance spectra.

Figure 3(b) presents the simulated reflectance spectra at normal incidence, where the side length *L* is varied from 120 nm to 220 nm while the value of *L'* is calculated based on Eq. (3) with *g'* being fixed at 12 nm. Here, we select the side length *L* to be 146 nm (*i.e. L'*=90 nm) such that photon absorption is enhanced by the 1st resonance at $\lambda_{pump}$ and the radiative decay rate is enhanced by the 2nd mode at $\lambda_{emission}$. Based on these dimension parameters, Fig. 3(c) presents the SEM image of the fabricated mix nanoantenna, with the measured and simulated angular-resolved reflectance spectra in Fig. 3(d). To understand the resonant optical modes, Fig. S17(a) presents the multipolar decomposition, where *ED* is the most dominating component at $\lambda_{pump}$ and $\lambda_{emission}$.

For the mix nanoantenna, the 1st contributing factor, enhanced absorption of pump laser, can be calculated based on the mode distribution at 532 nm in Fig. 4(a), considering the overlap between the enhanced optical field distribution and R6G molecules. The absorption enhancement factor is calculated to be ~25 fold. Moreover, FDTD simulation shows that c-Si antenna array absorbs 71.4% of the incident light power while R6G molecules absorb the remaining 24.8%, where monolayer R6G molecules with a thickness of 1.4 nm[52] is covering the exterior antenna surface based on the refractive index of R6G from the literature.[55]

The 2nd factor comes from the enhanced radiative decay rates, where the R6G emission lifetime from the mix antenna array is ~68 *p*s as shown in Fig. 4(b). In comparison to the flat substrate one, the total decay rate is enhanced by ~47 fold (*i.e.* Purcell factor $F_P$=47), which is due to the optical mode at $\lambda_{emission}$ (see Fig. S18). This radiative decay rate enhancement by the mix antenna is slightly better than the one by the square antenna in Fig. 2(b). Based on the simulated quantum yield on flat substrate (*QY*=~45.3%, see Fig. S9) and on the mix antenna (*QY'*=~50%, see Fig. S19), the enhanced radiative decay rate is estimated to be ~52 fold based on Eq. (2).

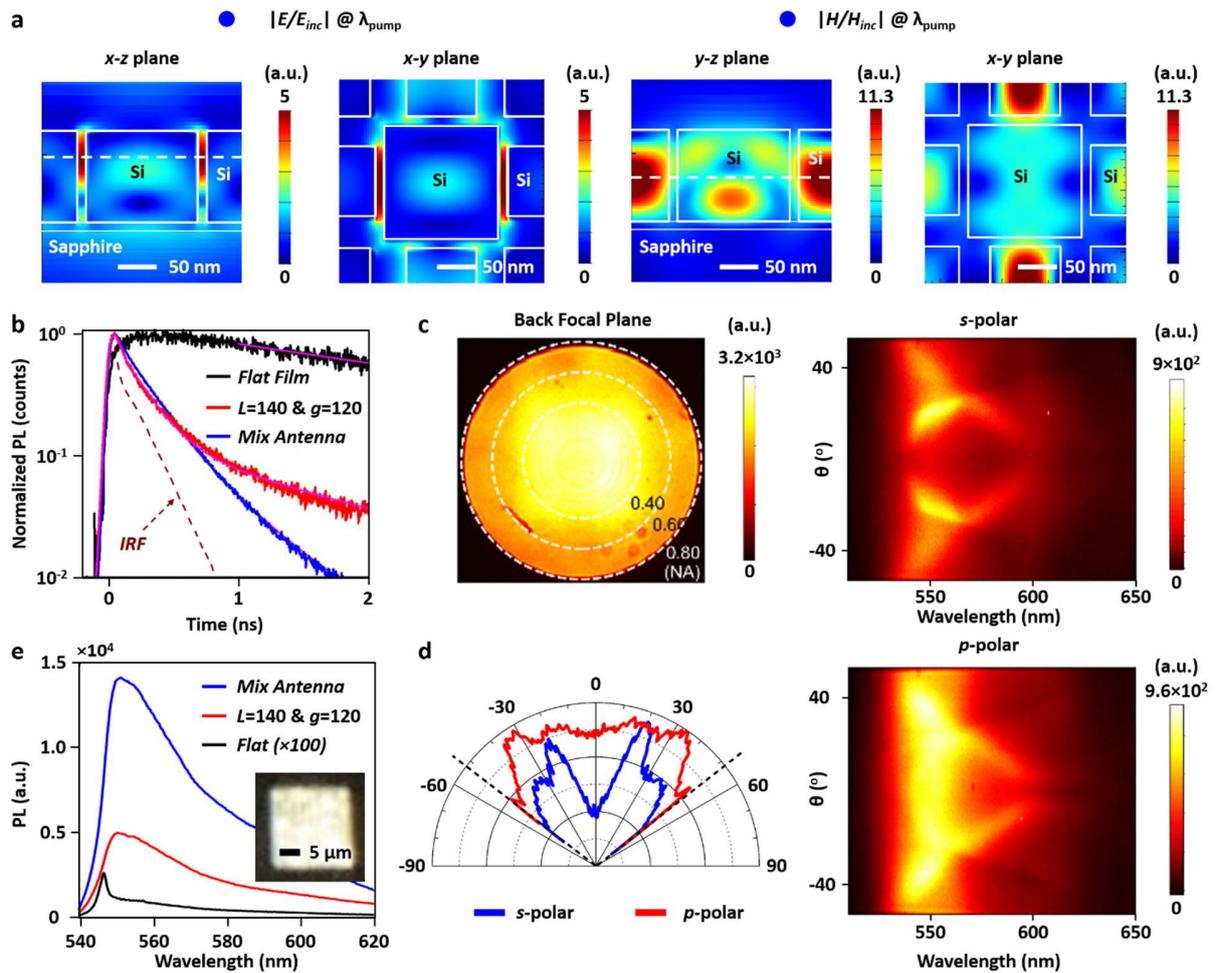

**Figure 4. Directional PL emission from mix c-Si nanoantenna array.** (a) Simulated field distributions ($|E/E_{inc}|$ and $|H/H_{inc}|$) at $\lambda_{pump}$ of 532 nm. (b) Enhanced radiative decay rate based on emission lifetime measurements. The emission lifetime was reduced from 3.2 ns (on flat substrate) to 68 ps (on mix antenna). (c) PL directionality characterization at BFP. (d) Polar plots of s- and p-polarized PL emission. (e) PL spectra as measured from the mix nanoantenna, square antenna (L=140 nm & g=120 nm) and flat region, by a 532 nm CW pump laser. The inset figure is the wide-field PL image of the mix antenna array with R6G molecules, and it is able to achieve an enhancement factor of ~1200 fold experimentally.

Moreover, the 3$^{rd}$ factor for the PL enhancement is directionality. Figure 4(c), Figure 4(d) and Fig. S20 present the directional emission characteristics of R6G molecules as deposited onto the mix nanoantenna. The enhanced PL directionality is calculated to be ~2.45 fold, as compared to the flat substrate case in Fig. S14. Therefore, the overall PL enhancement factor is estimated to be ~3185 fold based on Eq. (1). Experimentally, this mix nanoantenna array is able to achieve the PL (under unpolarized condition) enhancement factor of ~1200 fold as shown in Fig. 4(e) with a 532 nm pump laser, which is ~3× higher than the one as achieved by the simple square antenna. The increased enhancement mainly comes from the excellent absorption of the mix nanoantenna resonant at the pump wavelength.

Though the surface area of the patterned regions is larger than flat regions, the molecular coverage is not fully conformal as they were deposited using a spin coating method. Alternatively, one could choose to normalize the enhancement by considering only molecules within the trench, which could lead to an overestimate. To allow for consistent comparison with enhancement factors reported in literature,[35, 39] we simply use the planar area of the sample in our calculations. In addition, we investigated the spectral correlation between "maximum excitation rate" and "maximum extinction" of mix antenna array via FDTD simulations. Simulations consider a continuous film of R6G molecules with a thickness of 1.4 nm that coats the surface of the mix antenna array. The detailed simulation results are shown in Fig. S21. The peaks in the simulated excitation rate (a near-field effect) in Fig. S21(a) are aligned with the ones in the simulated extinction spectrum (a far-field measurement) in Fig. S21(b), with no observable wavelength shift between the local maxima of the two spectra. This result is unlike the case of plasmonic nanoantennas that behave as harmonic oscillators, where near-field effects are redshifted relative to far-field extinction spectra.[56] This discrepancy is probably due to a lower damping in the Si antenna system compared to plasmonic counterparts and could be a topic for further studies.

In conclusion, this paper presents a mix Si nanoantenna array capable of achieving trifecta fluorescence enhancements, *i.e.* through enhancements in absorption, radiative decay rates, and directionality. Monolayer R6G molecules deposited onto the designed mix nanoantenna array exhibited total PL enhancement of ~1200 fold and a Purcell factor of ~47, over measurements in the absence of nanoantennas. Careful analysis of our experiments and simulations show that this enhancement is attributed to the enhanced absorption of ~25 fold, enhanced radiative decay rates by ~52 fold and enhanced directionality by ~2.45 fold, with a quantum yield of ~50%. Such a high quantum yield on mix Si antenna provides a key advantage over the plasmonic counter partners, which are well known for the high non-radiative decay rates for quantum emitters being deposited onto the surface.[27] We believe that our work can lead to novel CMOS compatible nanoantenna platforms for enhancing weak fluorescence for biological and chemical applications,[43-45] as well as electrical excitation of quantum emitters.[57, 58]

## ASSOCIATED CONTENT

**Supporting Information**

The Supporting Information is available free of charge on the ACS Publications website.

Wood's Anomaly Calculation, fabrication processes, *n* and *k* values for the single crystalline Si film, scattering cross section plots, wide-field PL image, dependence of PL intensity on $\lambda_{pump}$, measured emission lifetime, relationship between lifetime, quantum yield and Purcell factor enhancement, simulated quantum yield on flat substrate, simulated quantum yield on the square nanoantenna array, angular-resolved PL spectra, mode distribution at $\lambda_{emission}$, simulated directionality on square antenna, simulated directionality on a flat sapphire substrate with a 10 nm thick Si film, PL from the nanoantenna and flat substrate region, evolution from "square" to "mix" array, multipolar decomposition for mix nanoantenna array, mode pattern at $\lambda_{emission}$, quantum yield calculation for mix nanoantenna array, simulated directionality for mix nanoantenna array, simulated spectra for the excitation rate and extinction of mix antenna array, methods section.

**AUTHOR INFORMATION**

**Corresponding Authors**

*E-mail: joel_yang@sutd.edu.sg.

*E-mail: Arseniy_Kuznetsov@imre.a-star.edu.sg.

*E-mail: dongz@imre.a-star.edu.sg.

**ORCID**

Zhaogang Dong: 0000-0002-0929-7723

Sergey Gorelik: 0000-0002-9215-8985

Ramón Paniagua-Dominguez: 0000-0001-7836-681X

Johnathan Yik: 000-0001-5872-5416

Jinfa Ho: 0000-0001-6884-4785

Emmanuel Lassalle: 0000-0002-0098-5159

Soroosh Daqiqeh Rezaei: 0000-0002-9807- 2074

Darren C. J. Neo: 0000-0001-6973-1117

Ping Bai: 0000-0002-7363-6478

Arseniy I. Kuznetsov: 0000-0002-7622-8939

Joel K. W. Yang: 0000-0003-3301-1040

**Author Contributions**

Z.D., A.I.K. and J.K.W.Y. conceived the concept, designed the experiments, and wrote the manuscript. Z.D. performed nanofabrication, and SEM imaging. S.G. did the optical characterizations. R.P.-D. performed the simulations for angular reflectance, multipolar decomposition, quantum yield, and Purcell factor. R.P.-D., J.Y., E.L. and P.B. performed the reciprocity calculations. J.H. and S.D.R. did the FDTD simulation. F.T. did the dry etching. D.C.J.N. measured the absorption of R6G molecules in solution. All authors analyzed the data, read and corrected the manuscript before submission.

**Notes**

The authors declare no competing financial interests.

**ACKNOWLEDGMENTS**

This work is supported by A*STAR SERC Pharos project (1527300025). Z.D. and J.K.W.Y. also acknowledges the support by A*STAR AME IRG (A20E5c0093) and A*STAR career development award.

**For Table of Contents (TOC) Use Only**

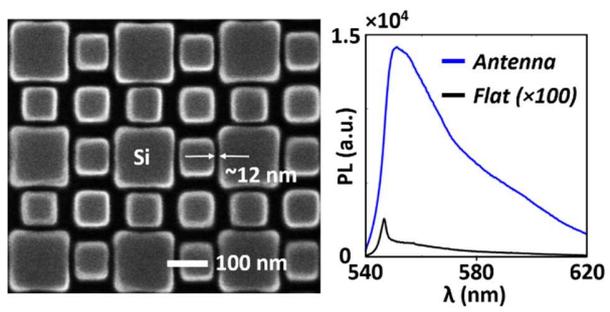

**Supporting Information for:**

**Silicon Nanoantenna Mix Arrays for a Trifecta of Quantum Emitter Enhancements**

Zhaogang Dong[1,*], Sergey Gorelik[1], Ramón Paniagua-Dominguez[1], Johnathan Yik[2], Jinfa Ho[1], Febiana Tjiptoharsono[1], Emmanuel Lassalle[1], Soroosh Daqiqeh Rezaei[3], Darren C. J. Neo[1], Ping Bai[2], Arseniy I. Kuznetsov[1,*], and Joel K. W. Yang[3,1,*]

[1]Institute of Materials Research and Engineering, A*STAR (Agency for Science, Technology and Research), 2 Fusionopolis Way, #08-03 Innovis, 138634 Singapore

[2]Institute of High Performance Computing, A*STAR (Agency for Science, Technology and Research), 1 Fusionopolis Way, #16-16 Connexis, 138632 Singapore

[3]Singapore University of Technology and Design, 8 Somapah Road, 487372, Singapore

*Correspondence and requests for materials should be addressed to J.K.W.Y. (email: joel_yang@sutd.edu.sg), A.I.K. (email: Arseniy_Kuznetsov@imre.a-star.edu.sg) and Z.D. (email: dongz@imre.a-star.edu.sg).

## S1. Wood's Anomaly Calculation.

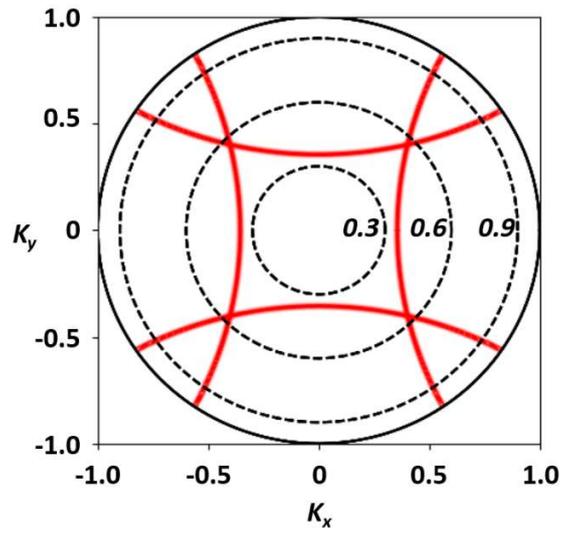

**Figure S1.** Calculations of Wood's Anomaly conditions for Si nanoantenna array with a pitch of 260 nm at the emission wavelength of 550 nm due to the sapphire substrate. The dashed lines indicate a numerical aperture (NA) of 0.30, 0.60 and 0.90 for reference.

## S2. Fabrication processes.

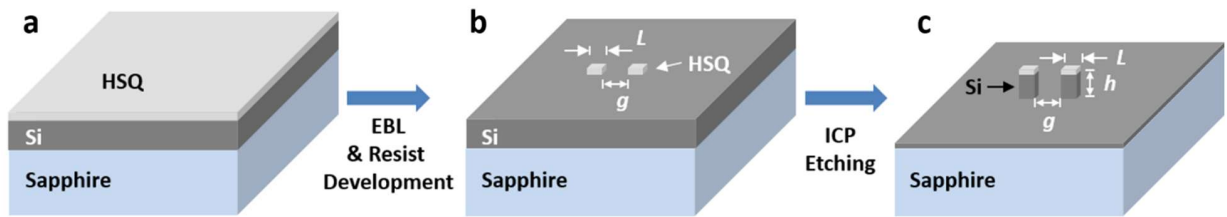

**Figure S2.** Fabrication process for the Si nanoantenna array. (a) Spin coating of hydrogen silsesquioxane (HSQ) resist onto the substrate. (b) E-beam exposure of HSQ resist followed by the salty development for 1 minute. The salty developer is NaOH/NaCl solution (1% wt./4% wt. in de-ionized water). (c) Inductively coupled plasma (ICP) dry etching of silicon nanostructures. The etching time is optimized such that there is 10 nm thick silicon left. $D$, $g$ and $h$ denote the diameter, gap size and height of the silicon nanopost antenna, respectively. After that, hydrofluoric (HF) acid (~10% in DI water) treatment for 1 minute is carried out to remove the residual HSQ resist.

**S3. n and k values of the single crystalline Si film.**

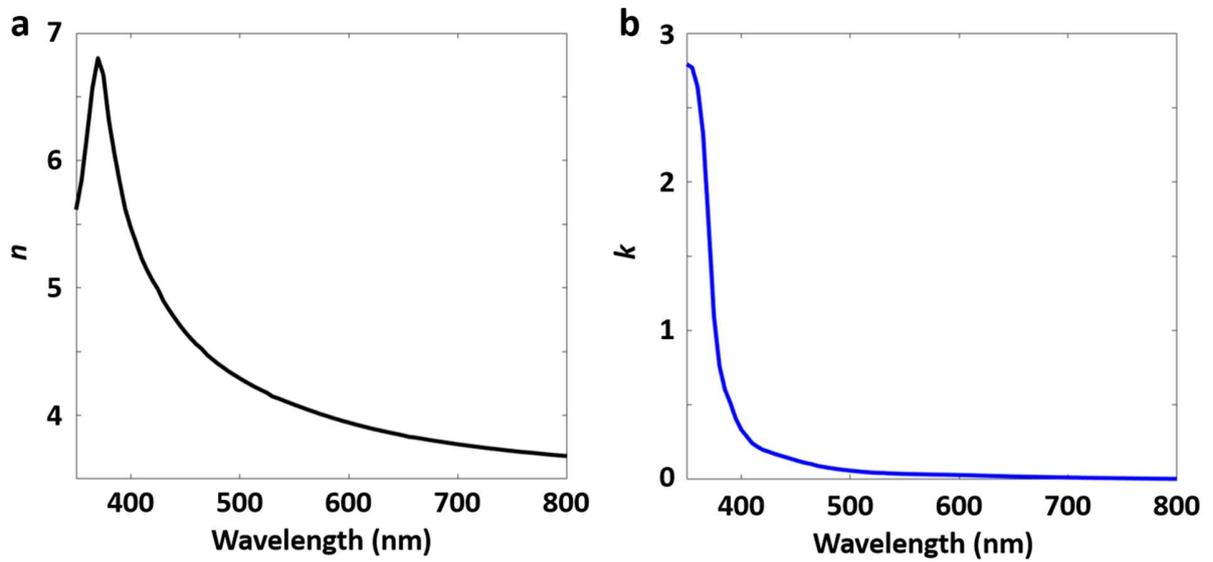

**Figure S3.** Measured *n* and *k* values of the single crystalline silicon film with a thickness of 130 nm as grown on sapphire substrate (Silicon Valley Microelectronics, Inc.).

## S4. Scattering cross section plots.

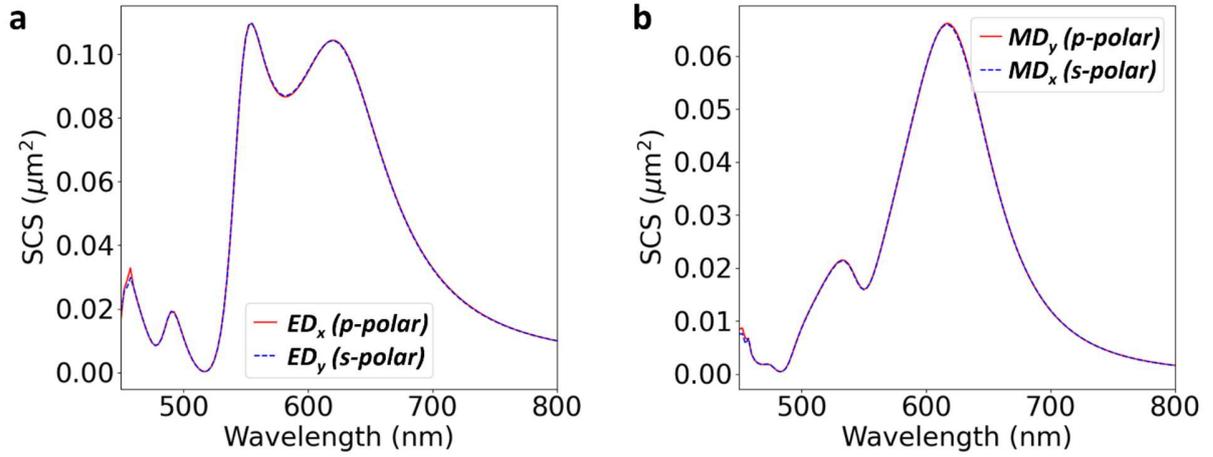

**Figure S4.** Scattering cross section (SCS) plots of the major multipolar decomposition components when θ=0 degree for the square silicon nanoantenna array in Figure 1. (a) $ED_x$ (@ *p*-polarized) and $ED_y$ (@ *s*-polarized) being identical when θ=0 degree. (b) $MD_x$ (@ *p*-polarized) and $MD_y$ (@ *s*-polarized) being identical when θ=0 degree.

**S5. Wide-field PL image.**

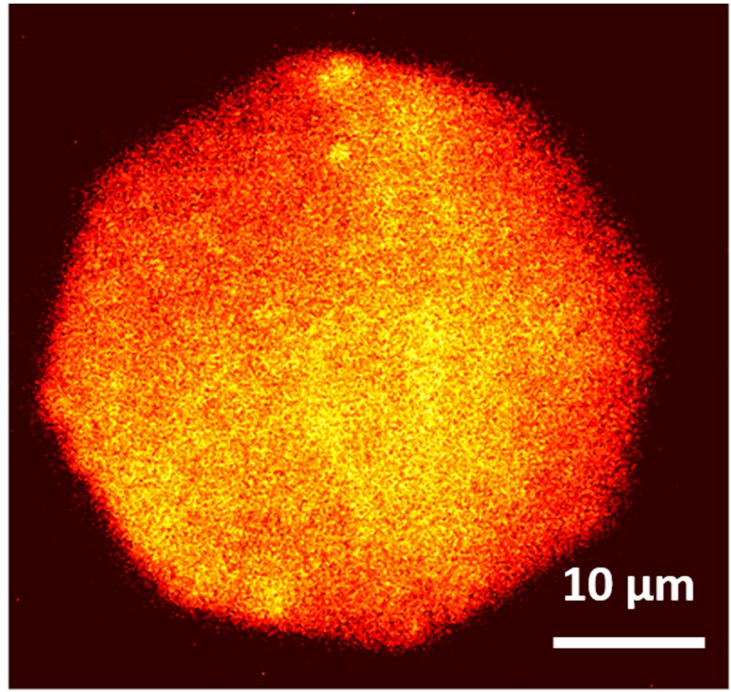

**Figure S5.** Wide-field PL image showing the uniformity of the R6G film deposited onto the Si nanoantenna array sample ($L$=140 nm & $g$=120 nm). This wide-field PL image was taken using a ×60 objective lens with a NA of 0.80. The octagon shape is due to the aperture of the microscope (not the patterned array).

## S6. Dependence of PL intensity on λ$_{pump}$.

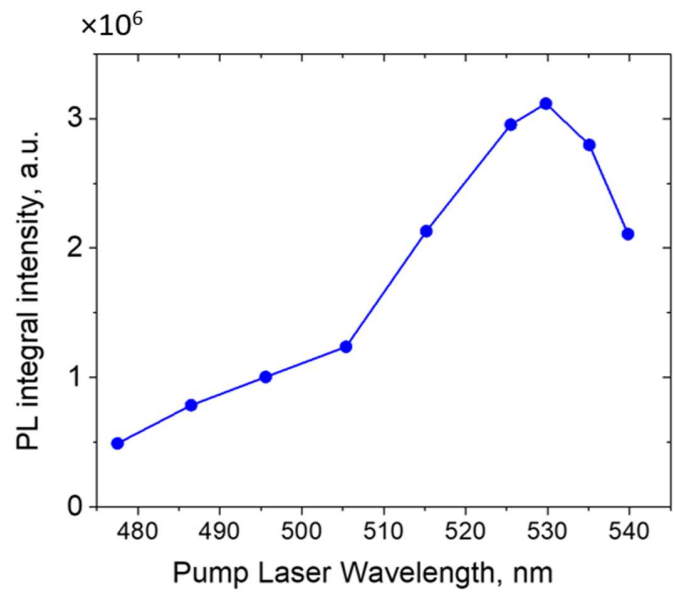

**Figure S6.** Dependence of integrated PL intensity on the pump laser wavelength. It shows that the most efficient pump laser wavelength is ~530 nm.

## S7. Measured emission lifetime.

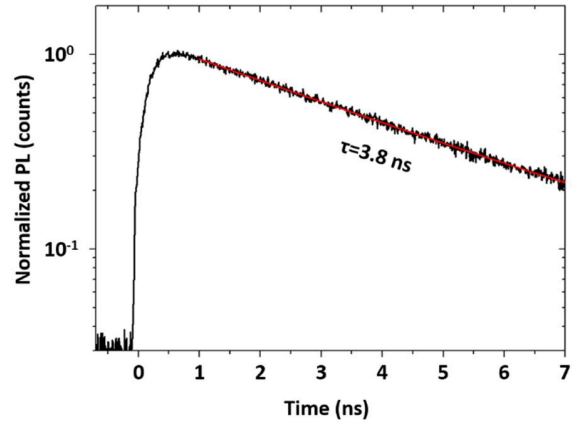

**Figure S7.** Measured Emission lifetime of R6G molecules in ethanol solution with a concentration of 1 μmol/L.

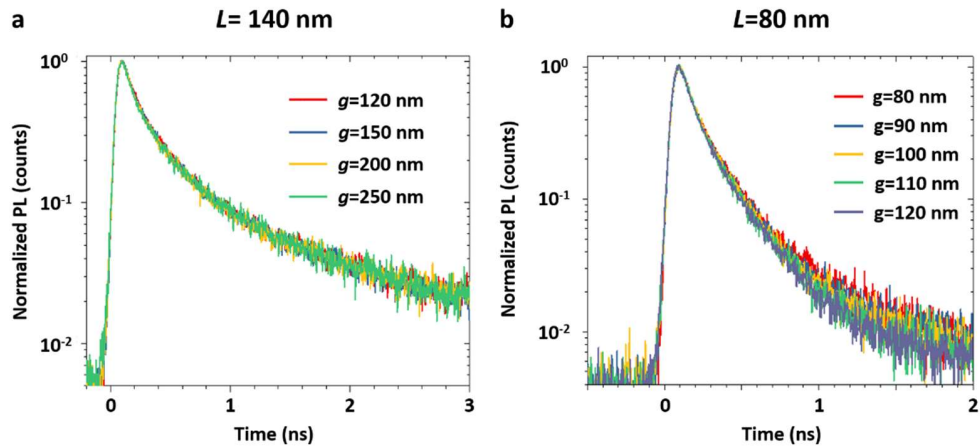

**Figure S8.** Emission lifetime of R6G molecule as deposited onto the square Si nanoantenna arrays. (a) Nanoantenna arrays with a side length *L* of 140 nm and different gaps g. (b) Nanoantenna arrays with a side length *L* of 80 nm and different gaps g. Table S1 presents a summary of the fitted emission lifetimes of R6G molecules deposited onto the respective Si nanoantenna arrays. The formula for fitting the PL decay profile is $A_1 e^{-\frac{t}{\tau_1}} + A_2 e^{-\frac{t}{\tau_2}}$. Table S1 presents a summary of the fitted emission lifetimes of R6G molecules deposited onto the respective Si nanoantenna arrays. Here, $R_1$ is defined as $A_1\tau_1/(A_1\tau_1+A_2\tau_2)$ and $R_2$ is defined as $A_2\tau_2/(A_1\tau_1+A_2\tau_2)$.

Table S1: Fitted emission lifetime components for the respective nanoantenna arrays.

| $L$ (nm) | $g$ (nm) | $A_1$ (a.u.) | $\tau_1$ (ns) | $A_2$ (a.u.) | $\tau_2$ (ns) | $A_1\tau_1$ | $A_2\tau_2$ | $R_1$ | $R_2$ |
|---|---|---|---|---|---|---|---|---|---|
| 130 | 120 | 41721 | 0.020 | 380 | 1.50 | 834.42 | 570 | 0.594 | 0.406 |
| 140 | 120 | 45145 | 0.022 | 580 | 1.63 | 993.19 | 945.4 | 0.512 | 0.488 |
| 150 | 120 | 37465 | 0.022 | 660 | 1.47 | 824.23 | 970.2 | 0.459 | 0.541 |
| 160 | 120 | 51207 | 0.021 | 800 | 1.61 | 1075.35 | 1288 | 0.455 | 0.545 |
| 170 | 120 | 50721 | 0.024 | 1087 | 1.49 | 1217.3 | 1619.5 | 0.429 | 0.571 |
| 180 | 120 | 26350 | 0.024 | 643 | 1.47 | 632.4 | 945.21 | 0.400 | 0.600 |
| 200 | 120 | 15804 | 0.052 | 1217 | 1.30 | 821.81 | 1582 | 0.342 | 0.658 |
| 210 | 120 | 21900 | 0.032 | 1393 | 1.45 | 700.80 | 2019.85 | 0.258 | 0.742 |
| 220 | 120 | 15546 | 0.040 | 1366 | 1.43 | 621.84 | 1953.38 | 0.241 | 0.759 |
| Flat film, on the substrate used || 0 | 0 | 331 | 3.0 | 0 | 926.80 | 0 | 1 |
| Flat film, on glass || 0 | 0 | 1 | 3.2 | 0 | 3.2 | 0 | 1 |
| Ethanol solution, 1 μM || 0 | 0 | 1 | 3.8 | 0 | 3.8 | 0 | 1 |

## S8. Relationship between lifetime, quantum yield and Purcell factor enhancement.

The enhanced radiative decay rate can be attributed to the Purcell factor $F_P$ defined as:[1,2]

$$F_P = \frac{3}{4\pi^2}\left(\frac{Q}{V_{mode}}\right)\left(\frac{\lambda}{2n}\right)^3, \tag{S1}$$

where $Q$ denotes the quality factor of the nanoantenna array resonance; $V_{mode}$ denotes the mode volume; $\lambda$ is the emission wavelength; and $n$ is the refractive index of the environment. This equation shows that the Purcell factor is proportional to $Q$ and inversely proportional to $V_{mode}$.[1]

The total decay rate on the flat substrate region (without the nanoantenna effect) is denoted as $\gamma_{total}$. It is a sum of the radiative decay rate $\gamma_r$ and the non-radiative decay rate $\gamma_{nr}$ as given by:

$$\gamma_{total} = \gamma_r + \gamma_{nr}. \tag{S2}$$

Furthermore, the quantum yield $QY$ for the fluorescent molecules as deposited onto the flat substrate is related to the radiative decay rate $\gamma_r$ and the non-radiative decay rate $\gamma_{nr}$ by the following equation:

$$QY = \frac{\gamma_r}{\gamma_r + \gamma_{nr}}. \tag{S3}$$

The total decay rate with the nanoantennas is denoted as $\gamma'_{total}$ and is given by:

$$\gamma'_{total} = \gamma_r' + \gamma_{nr}', \tag{S4}$$

where $\gamma_r'$ denotes the radiative decay rate with nanoantenna and $\gamma_{nr}'$ denotes the non-radiative decay rate with nanoantenna. The quantum yield $QY'$ for the fluorescent molecules as deposited onto the nanoantennas is related to the radiative decay rate and the non-radiative decay rate by the following equation:

$$QY' = \frac{\gamma_r'}{\gamma_r' + \gamma_{nr}'}. \tag{S5}$$

Due to the Purcell effect of the nanoantennas, the total decay rate is modified and it is given by:

$$\gamma'_{total} = \gamma_{total} \times F_P, \tag{S6}$$

where $F_P$ can be determined experimentally based on the measured total decay rate at the flat substrate region $\gamma_{total}$, as well as the measured total decay rate with at the nanoantenna region $\gamma'_{total}$. From the above formulas, the Purcell factor $F_P$ is related to the radiative decay rate and the quantum yield by the following equation:

$$\frac{\gamma'_r}{\gamma_r} = F_P \times \frac{QY}{QY}. \tag{S7}$$

## S9. Simulated quantum yield on flat substrate.

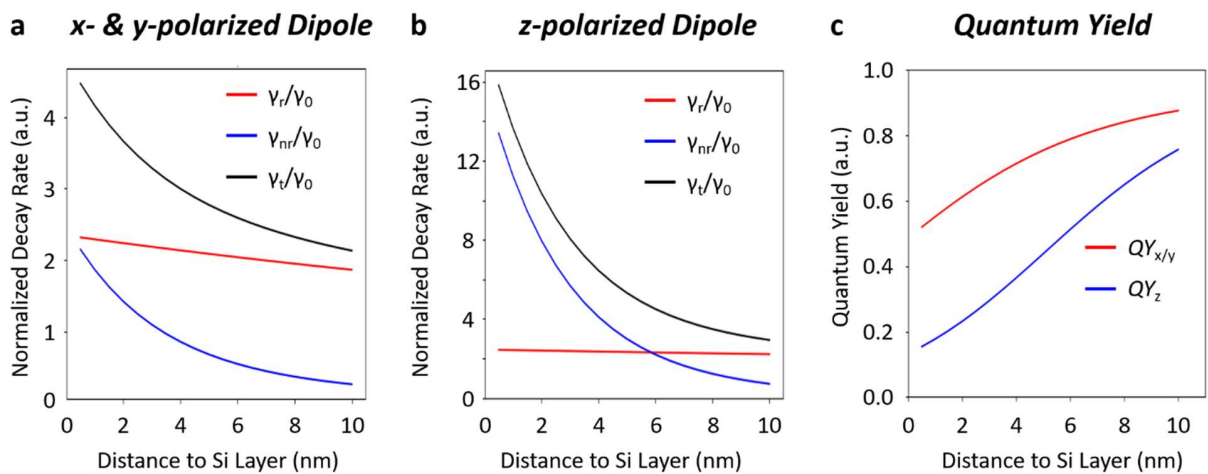

**Figure S9. Simulation of the quantum yield of dipole sources on a flat sapphire substrate with a 10 nm thick Si film at the emission wavelength of 550 nm.** (a) Normalized decay rates of an *x*-/*y*-polarized dipole with respect to the distance between the dipole source and the top surface of Si film. The normalization of the decay rate is done with respect to the dipole source in free-space. (b) Normalized decay rates of a *z*-polarized dipole with respect to the distance between dipole source and the top surface of Si film. (c) Simulated quantum yields for a *x*-/*y*- dipole source and a *z*-polarized dipole source. Considering the random dipole orientation and the monolayer of R6G molecule with a thickness of 1.4 nm,[3] the averaged quantum yield *QY* is ~45.3%.

**S10. Simulated quantum yield on the square nanoantenna array.**

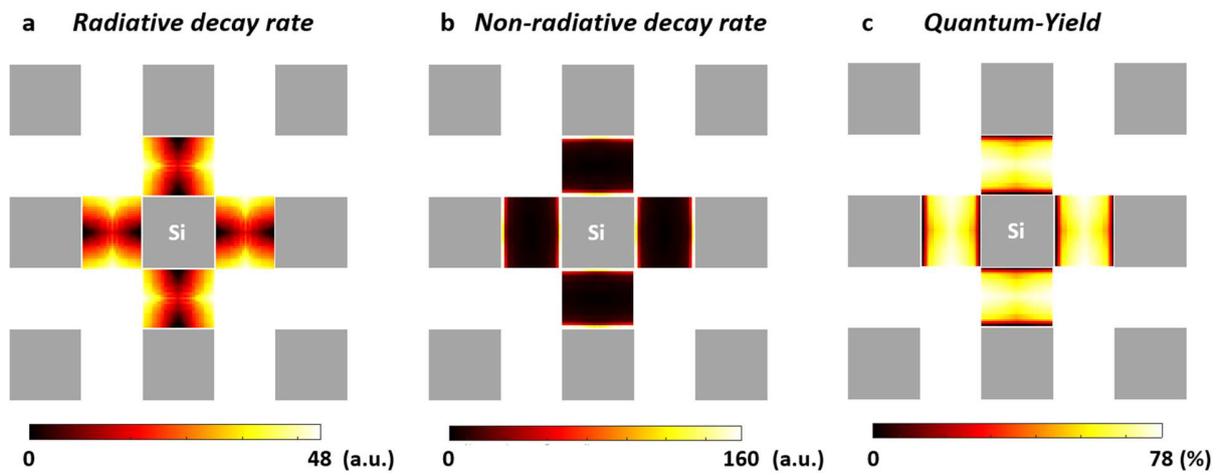

**Figure S10. Simulated quantum yield of the square nanoantenna array ($L$=140 nm & $g$=120 nm) at the emission wavelength of 550 nm (considering random dipole orientations).** (a) Radiative decay rate. (b) Non-radiative decay rate. (c) Quantum yield. It shows that the simulated quantum yield $QY'$ is around ~40% for the dipole, which is placed 1.4 nm above the square nanoantenna array surface.

## S11. Angular-resolved PL spectra.

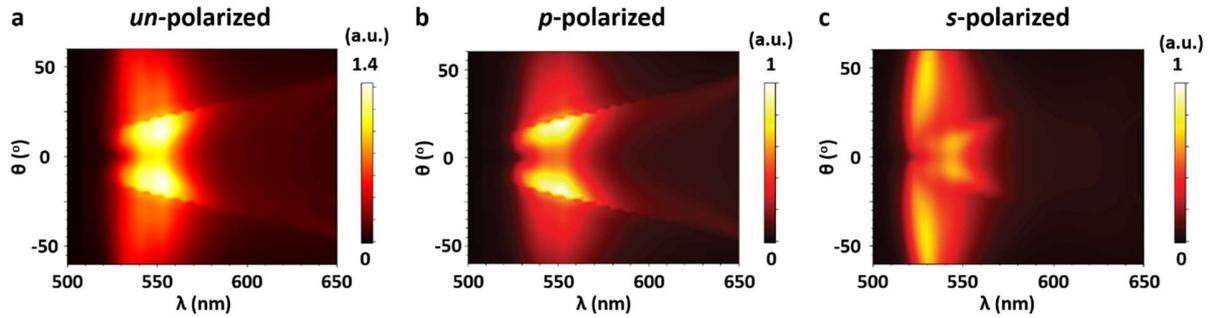

**Figure S11.** (a)-(c) Simulated angular-resolved PL spectra under respective polarization conditions. The simulated PL plots are normalized to the maxima of *p*-polarized emission. Both experimental (*i.e.*, Figure 2(c)) and simulation results reveal that at the small solid angle region, the overall PL is dominated by the *p*-polarized emission. In other words, the *p*-polarized emission possesses good directionality, as shown in Fig. 2(c), with peak emission lobes at ~12° with respect to the normal direction. Such a strong directionality originates from the coupling of the R6G emission to the hybrid Mie-diffraction mode, which is due to the optical mode of the nanoantenna array as shown in Fig. S12. In comparison, R6G coupled to the *s*-polarized mode emits over a large angular range, as localized Mie resonances excited in this case do not couple efficiently to the diffraction (or lattice) modes.

**S12. Mode distribution at λ_emission.**

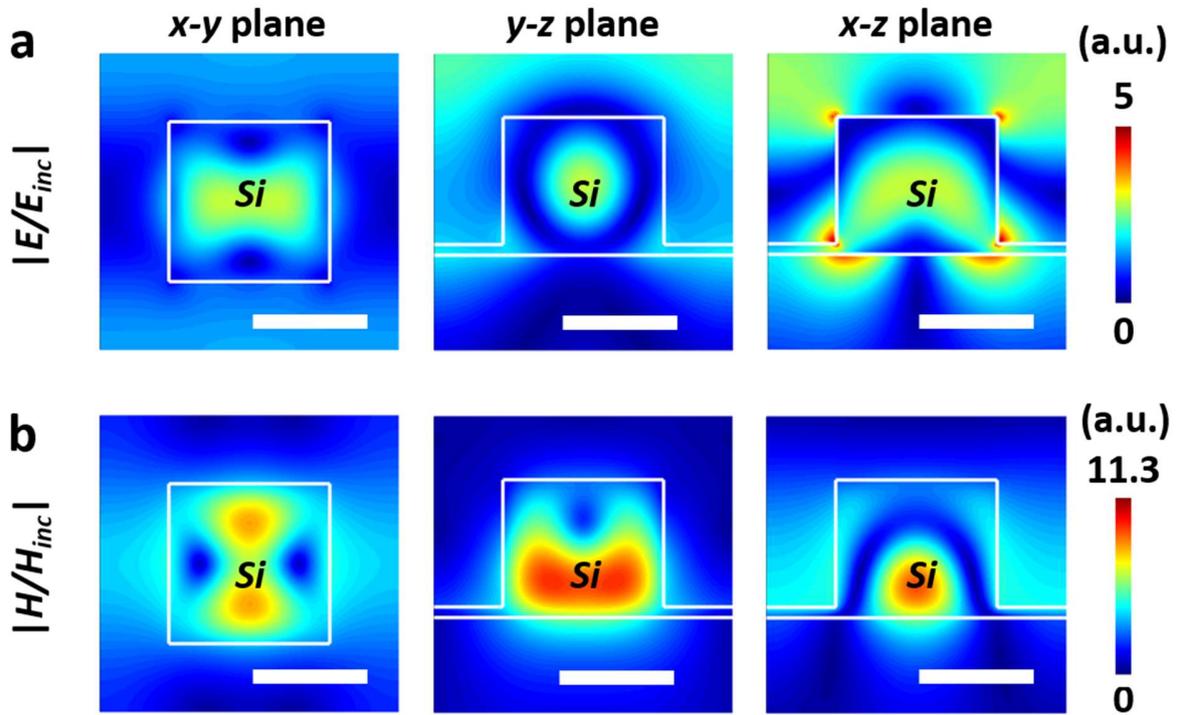

**Figure S12.** (a)-(b) Normalized field distributions (|*E/E$_{inc}$*| and |*H/H$_{inc}$*|) at the emission wavelength of 550 nm for the square nanoantenna array. The scale bar denotes 100 nm. *E$_{inc}$* and *H$_{inc}$* denote the amplitude of electric field and magnetic field in free-space.

**S13. Simulated directionality on square antenna.**

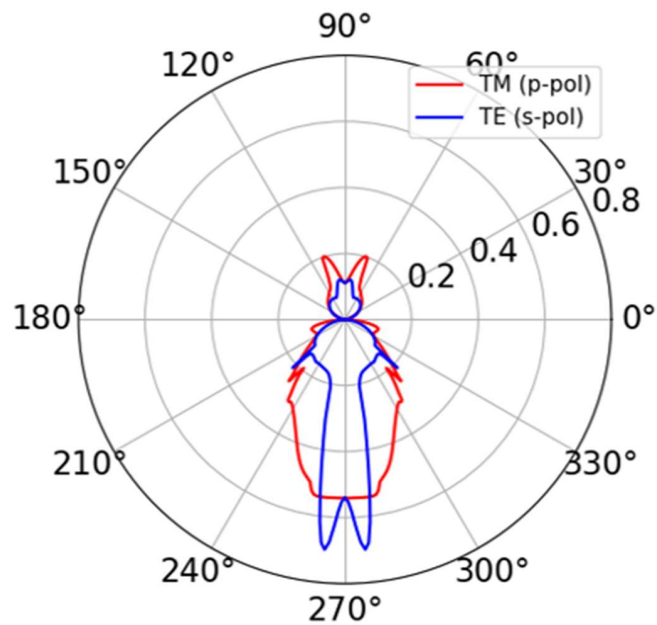

**Figure S13. Simulated emission directionality of the square Si nanoantenna array ($L$=140 nm and $g$=120 nm) obtained by the reciprocity theory calculations (see Methods section for details).** Averaging the 3 dipole orientations and different polarizations leads to the percentage of the emitting power into air of 24.7% and the rest 75.3% are emitting into the substrate. The current sapphire substrate is not polished on the back side, and thus we cannot collect the PL signal from the substrate side.

**S14. Simulated directionality on a flat sapphire substrate with a 10 nm thick Si film.**

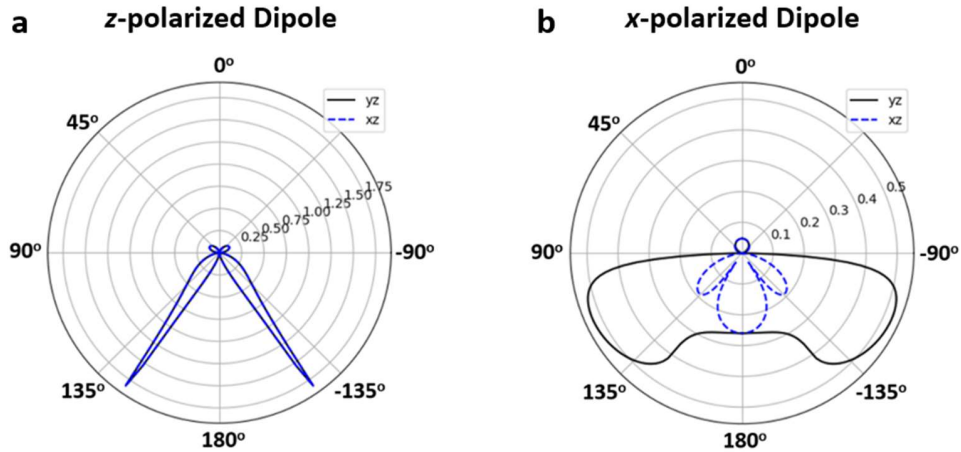

**Figure S14. Simulated emission directionality of dipole sources placed on top of a flat sapphire substrate with a 10 nm thick Si film.** (a) *z*-polarized dipole. When the dipole source is *z*-polarized, 19.5% and 80.5% of the total energy is emitted towards the air region and the substrate respectively. (b) *x*-polarized and *y*-polarized dipoles. When the dipole source is *x*-polarized or *y*-polarized, 8.6% and 91.4% of the total energy is emitted towards the air region and the substrate region, respectively. Considering random dipole orientations, the percentages of power emitting towards the air region and the substrate region are 12.2% and 87.8%, respectively. Comparing the flat substrate case and the square nanoantenna array in Fig. S13, the enhanced directionality of the PL emitting into the air region is calculated to be ~2 fold.

**S15. PL from the nanoantenna and flat substrate region.**

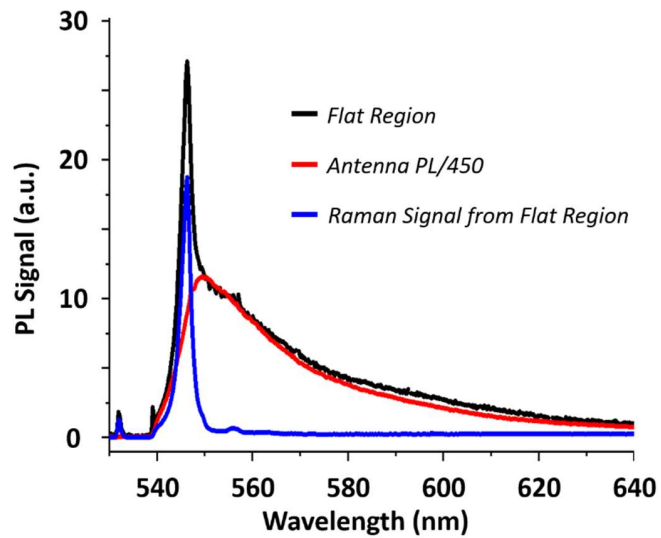

**Figure S15. Comparison of the PL emissions from the nanoantenna region and the flat substrate region.** The black curve presents the measured PL emission spectrum from the flat region, and it is then fitted with one sharp PL component and one broad PL component. The sharp PL component represents the Raman feature from the flat substrate region with a Raman shift of ~482 cm$^{-1}$ due to sapphire.[4] The broad PL component is originated from the R6G PL emission, which is 1/450 of the one from the square nanoantenna region.

**S16. Evolution from "square" to "mix" array.**

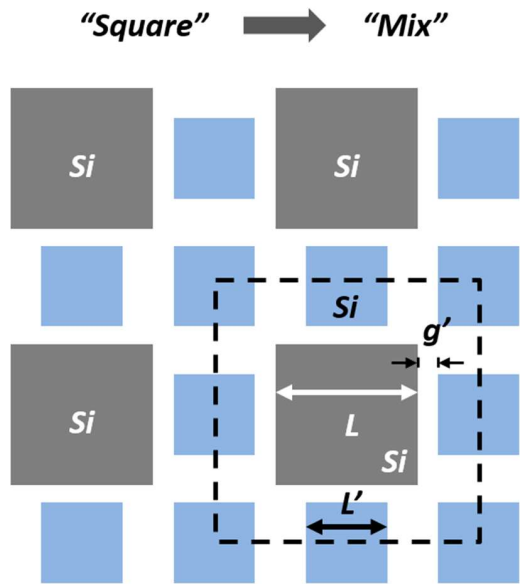

**Figure S16. Evolution from the "square" nanoantenna array to "mix" nanoantenna array.** A mix antenna design could be achieved by inserting small square elements into the square nanoantenna array. The effective gaps of the nanoantenna are getting smaller. First, smaller gaps are desired to enhance the pump laser absorption, because the optical intensity in the gap region is increasing exponentially with respect to the reduced gap size.[5] These smaller square elements can also lead to faster radiative decay rates as shown in Fig. 2(b).

## S17. Multipolar decomposition for mix nanoantenna array.

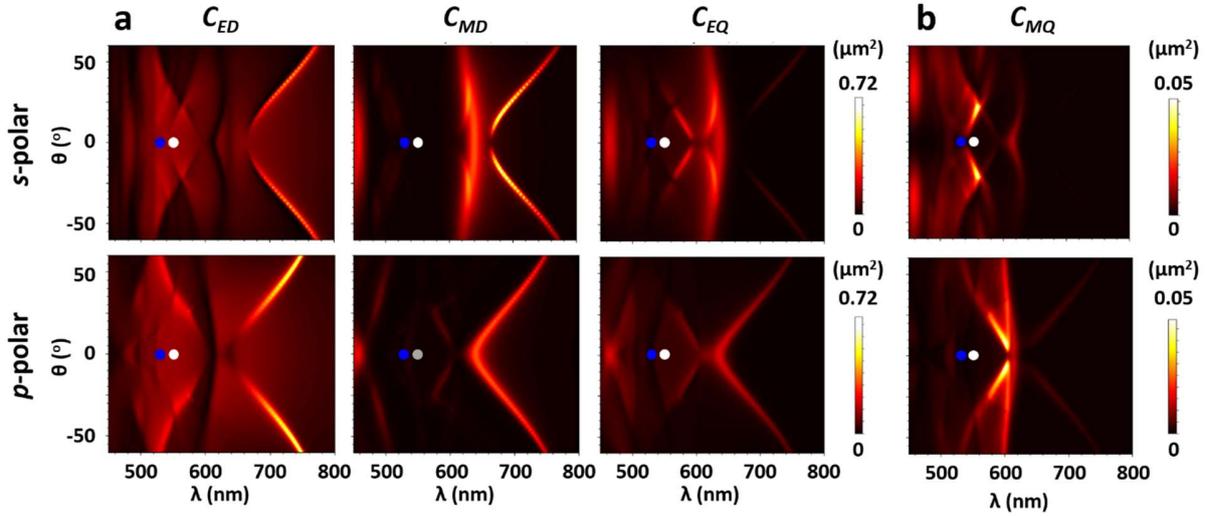

**Figure S17. Multipolar decomposition analysis for the mix nanoantenna array under both *s*- and *p*-polarized incidence conditions.** (a) Electric dipole (*ED*), magnetic dipole (*MD*), and electric quadrupole (*EQ*) as excited in the mix nanoantenna array. (b) Magnetic quadrupole (MQ) as excited in the mix nanoantenna array. It shows that the amplitude of MQ is one order of magnitude weaker than electric dipole (*ED*), magnetic dipole (*MD*) and electric quadrupole (*EQ*). The blue color dots and the white color dots are denoting the pumping wavelength $\lambda_{pump}$ and the emission wavelength $\lambda_{emission}$, respectively.

## S18. Mode pattern at λ$_{emission}$.

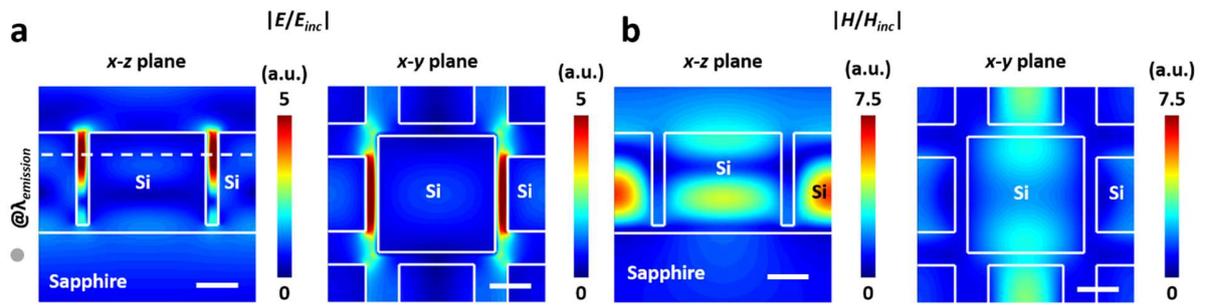

**Figure S18. Normalized field distributions around the mix nanoantenna at the emission wavelength of ~550 nm.** (a) |*E*/*E$_{inc}$*|. (b) |*H*/*H$_{inc}$*|. The scale bar denotes 50 nm. ***E**$_{inc}$* and ***H**$_{inc}$* denote the amplitude of electric field and magnetic field in free-space.

**S19. Quantum yield calculation for mix nanoantenna array.**

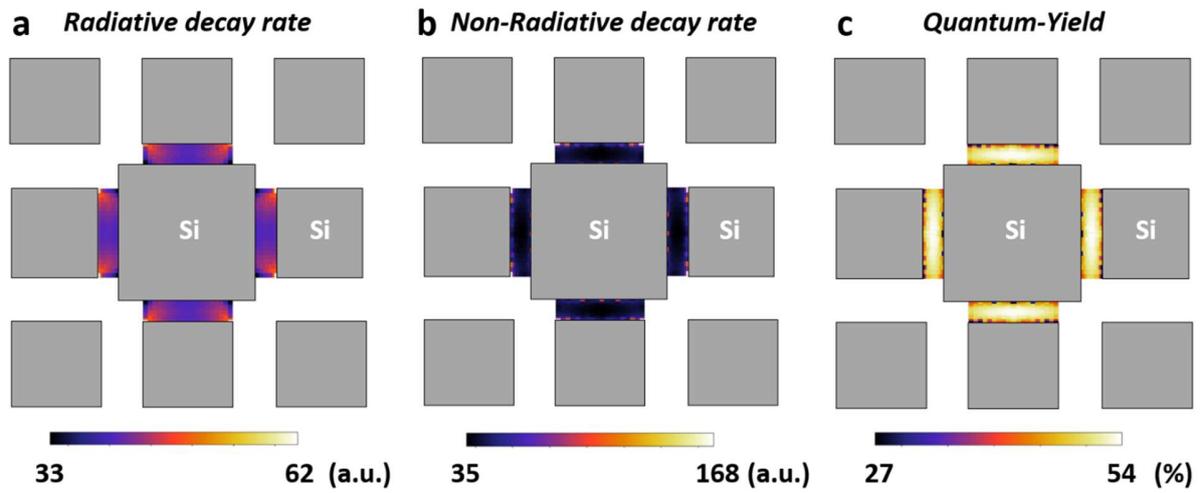

**Figure S19. Simulated quantum yield of the mix antenna array at the emission wavelength of 550 nm (considering random dipole orientations).** (a) Radiative decay rate. (b) Non-radiative decay rate. (c) Quantum yield. It shows that the simulated quantum yield is around ~50% for the dipole, which is placed ~1.4 nm above the surface of the mix nanoantenna array.

**S20. Simulated directionality for mix nanoantenna array.**

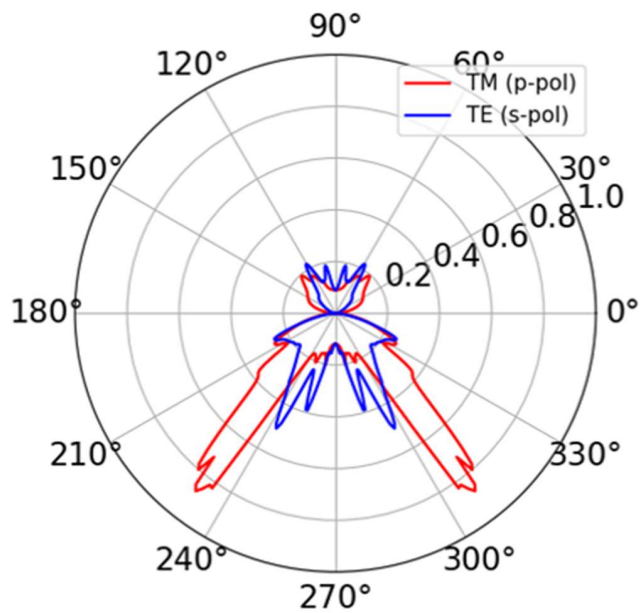

**Figure S20. Simulated emission directionality of the mix antenna array at the emission wavelength of 550 nm via reciprocal calculations for both *s*-polarized and *p*-polarized emission.** By averaging over the 3 dipole orientations and over different polarizations, the power portion emitting into air and into substrate is 29.9% and 70.1%, respectively. As compared to the flat substrate case in Fig. S14, the enhanced directionality of PL emitting into the air region is calculated to be ~2.45 fold.

**S21. Simulated spectra for the excitation rate and extinction of mix antenna array.**

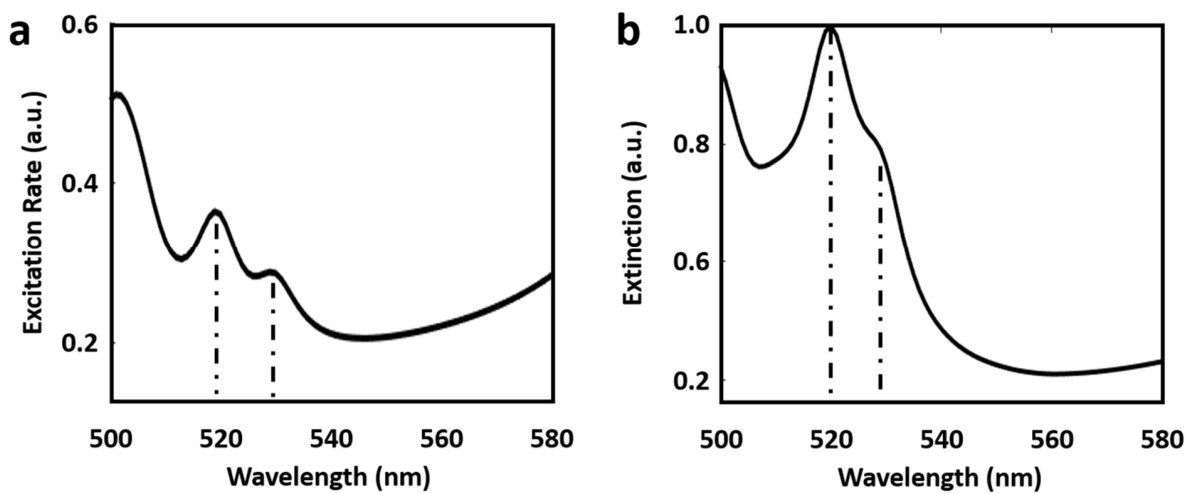

**Figure S21. Simulated spectra for the excitation rate and extinction of mix antenna array coated by a continuous film of R6G molecules with a thickness of 1.4 nm.** (a) Simulated excitation rate. This excitation rate is calculated based on the integrated sum of the near-field intensity $|E|^2$ within the R6G film. (b) Simulated extinction spectrum. The peaks in the simulated excitation rate in (a) are well aligned with the ones in the simulated extinction spectrum in (b). There is no obvious wavelength shift observed between the maxima of "excitation rate" with respect to maximum "extinction efficiency".

**S22. Methods section.**

**Fabrication of c-Si Nanoantenna Arrays.** Hydrogen silsesquioxane (HSQ) etching mask was fabricated on the 130 nm thick single crystalline Si (c-Si) as grown on sapphire substrate (Si Valley Microelectronics, Inc.). The back side of sapphire substrate is not polished. HSQ resist (Dow Corning XR-1541-002) was first spin coated onto a cleaned substrate at 5k round-per-minute (rpm) to obtain an HSQ thickness of ~30 nm.[6] Electron beam exposure was then carried out with an electron acceleration voltage of 100 keV (EBL, Elionix ELS-7000), beam current of 200 pA, and an exposure dose of ~12 mC/cm$^2$. The sample was then developed by NaOH/NaCl salty solution (1% wt./4% wt. in de-ionized water) for 60 seconds and then immersed in de-ionized water for 60 seconds to stop the development. The sample was immediately rinsed by acetone, isopropanol alcohol (IPA), and dried by a continuous flow of nitrogen gas. Si etching was then carried out by inductively-coupled-plasma (ICP, Oxford Instruments Plasmalab System 100),[6] with a DC power of 100 watts, coil power of 500 watts, $Cl_2$ with a flow rate of 22 sccm (standard-cubic-centimeters-per-minute), under a process pressure of 10 mTorr, and temperature of 6 °C.

**Scanning Electron Microscope (SEM) Characterizations.** SEM images were taken at an acceleration voltage of 10 keV with the SEM model number of Elionix, ESM-9000.

**Photoluminescence (PL) Characterizations.** PL emission lifetime measurements, PL back focal plane (BFP) measurements and PL intensity enhancement measurements were employed to investigate the Purcell effect, directionality and total PL enhancement. For the PL emission lifetime measurements, a 470 nm pulsed laser was used as an excitation source; for PL BFP imaging, a 488 nm CW laser was used as PL excitation source. In general, the characteristics of PL decay rate and PL directionality at the emission wavelength (*i.e.* 550 nm) do not depend on the excitation light wavelength. However, a 532 nm excitation laser was specifically used for the PL enhancement measurements because it is close to antenna's Mie resonances, which are able to enhance the absorption of pump laser energy. The detailed descriptions of PL enhancement measurements, back focal plane (BFP) characterizations and emission lifetime measurements are shown below.

**PL Enhancement Measurements.** To measure PL enhancement factor, a 532 nm CW laser was used as the excitation light source. Andor Sr-303i spectrograph equipped with Newton CCD was used as a detector coupled to Nikon inverted microscope (Nikon Ti-U) using a wide field EPI configuration. Samples were placed onto the object plane of the microscope, and antenna arrays with R6G film were facing the objective. Excitation light was centered at the antenna array surface. The excitation beam size on the sample was controlled by the field aperture of the microscope and the objective lens magnification. The objective lens used here is 150× with a NA of 0.90.

**Back Focal Plane (BFP) Characterization.** Angle-resolved reflectance spectra and PL directionality were studied using angular resolved BFP spectroscopy.[7] Briefly, an inverted optical microscope (Nikon Ti-U) was coupled to a spectrograph (Andor SR-303i) equipped with an EMCCD detector (Andor Newton). For measurements of angle-resolved reflectance spectra, light from a halogen lamp passing through a microscope field aperture was focused onto the sample surface using an objective lens (60×, NA of 0.8). Reflectance of an aluminum mirror was used as a reference. To measure angle-resolved PL spectra, a 488 nm excitation CW laser was focused onto the sample surface with a spot size of ~10 μm and a power of ~20 μW. The reflected PL emission was then collected using the same objective. The BFP of the objective was imaged onto the spectrograph entrance slit with a width of 100 μm. The detectable angle range was ±53° and ±72° for the 60× NA=0.8 and 100× NA=0.90 objective lens respectively. For reflectance measurements, the polarization of incident and reflected light was varied simultaneously with respect to spectrograph entrance slit using two polarizers in parallel configuration. For PL measurements, only the polarization of PL emission was varied. *P*-polarization corresponds to the polarization at the entrance slit being parallel to the slit, while *s*-polarization corresponds to that perpendicular to the slit. PL directionality was measured in BFP with the same setup configuration, where the spectrograph entrance slit was fully open together with the 0-order diffraction grating.

**Emission Lifetime Characterizations.** Time-resolved PL was studied using a Picoquant Microtime 200 TCSPC system coupled with an Olympus microscope in confocal configuration. A 470 nm pulsed laser, pulse duration 70 *p*s, 40 MHz repetition rate, average power about 1 µW was used to excite PL. The laser light was focused onto the sample surface using a 20x objective lens. Spectrally integrated PL in 500-600 nm range was collected using a Si single photon avalanche photodiode. The instrument response function was ~250 *p*s, where the PL decay curves were recorded and analyzed using a Picoharp software. The instrument response function (IRF) was recorded using excitation light scattered from the sample, where the IRF was measured to be ~250 ps. The measured IRF was then used for the multi-exponential reconvolution fit of the PL decay profiles and the fitting was carried out using the Picoharp software.[8]

**Numerical Simulations.** Finite-difference time-domain (FDTD) simulations of the reflectance spectra at normal incidence and electric field distribution were carried out using Lumerical FDTD Solutions. Periodic boundary conditions were used with the incident optical field being *x*-polarized. The refractive index of the c-Si film as grown on sapphire was taken from the ellipsometer measurement as shown in Fig. S2. The angular reflectance spectra and the multipolar decomposition, the details of which could be found in the Ref[9], where computed using finite-element method (FEM) simulations carried out in Comsol Multiphysics. Two periodic ports (top and bottom of the array) were used for excitation and reflection and transmission calculations (with corresponding diffraction orders), while Bloch boundary conditions were used in the transverse directions. Reciprocity theory calculations of the directional emission of R6G molecules as deposited on the antennas were performed using the same setup but the field is integrated in the region occupied by the R6G molecules. At each incidence angle, the total electric energy $\sim|E|^2$ is integrated at the surface of the array to determine the emission strength towards that specific direction via reciprocity. By varying the incidence angle from 0 degree to 360 degree, we are able to obtain the enhancement angle in all directions.